\documentclass[jap,aip,preprint]{revtex4-2}

\usepackage{graphicx}  
\usepackage{dcolumn}   
\usepackage{bm}        
\usepackage{amssymb}
\usepackage[english]{babel}
\usepackage{xcolor}

\begin{document}


\title{Pressure driven re-entrant magnetoelectric transition in honeycomb  $Fe_{4}Nb_{2}O_{9}$ }

\author{Mrinmay Sahu}
\affiliation {National Centre for High Pressure Studies, Department of Physical Sciences, Indian Institute of Science Education and Research Kolkata, Mohanpur Campus, Mohanpur – 741246, Nadia, West Bengal, India.}

\author{Bishnupada Ghosh}
\affiliation {National Centre for High Pressure Studies, Department of Physical Sciences, Indian Institute of Science Education and Research Kolkata, Mohanpur Campus, Mohanpur – 741246, Nadia, West Bengal, India.}

\author{Rajesh Jana}
\affiliation {Beijing National Laboratory for Condensed Matter Physics and Institute of Physics, Chinese Academy of Sciences, Beijing 100190, China}

\author{Jinguang Cheng}
\affiliation {Beijing National Laboratory for Condensed Matter Physics and Institute of Physics, Chinese Academy of Sciences, Beijing 100190, China}

\author{Goutam Dev Mukherjee}
\affiliation {National Centre for High Pressure Studies, Department of Physical Sciences, Indian Institute of Science Education and Research Kolkata, Mohanpur Campus, Mohanpur – 741246, Nadia, West Bengal, India.}
\email [Corresponding Author:]{ goutamdev@iiserkol.ac.in}

\date{\today}

\begin{abstract}

A detailed high pressure investigation is carried out on $Fe_4Nb_2O_9$ using angle resolved x-ray diffraction and Raman spectroscopy measurements. We find a structural transition from the ambient trigonal phase to a monoclinic phase above 8.8 GPa. The structural transition is assumed to be driven by a large distortion of $Nb-O_6$ octahedra as seen from x-ray diffraction analysis and a large pressure dependence of $Nb-O_6$ octahedra breathing Raman mode. Anomalous behaviour of Raman modes and increase in the phonon life time at the phase transition pressure indicate a trigonal paramagnetic  phase to a monoclinic  antiferromagnetic state above 8.8 GPa. Decrease in the diffusive scattering rate of low frequency electronic contribution contradicts the results of decrease in intensity of high frequency electronic response and excludes the phenomenon of insulator to metal transition. Instead the enhancement of the intensity of the Raman modes till about 8.8 GPa indicate a large change in ferroelectric polarization of the sample indicating pressure induced re-entrant magentoelectric effect in $Fe_4Nb_2O_9$.


\end{abstract}

\maketitle

\section{Introduction} 
Recent developments in information technology are linked to various devices, such as, multistage storage memory devices, sensors, spintronic devices etc., which utilize electric and magnetic properties of materials. The strongly correlated systems, which exhibit mutual coupling of multiple order parameters have attracted widespread research interest due to their innovative technological applications \cite{maignan2018fe}. Multiferroic materials hold great application possibilities for the development of electronic devices because of the coexistence of ferroelectric and magnetic orders and their outstanding mutual coupling \cite{maignan2018fe,cao2017single,deng2019large,yin2016colossal}. This can lead to significant magneto-electric (ME) coupling in multiferroic materials, in which, the magnetic information can be controlled by an electric field or the electric polarization by external magnetic field
\cite{rivera2009short,schmid1994multi}. There are spin-driven (type-II) multiferroics in which specific magnetic ordering is responsible for the inversion symmetry breaking. $CuO$, $GdFeO_3$, $TbMnO_3$ etc belong to type-II multiferroics and they exhibit ME coupling \cite{agyei1990linear,tokunaga2009composite,jana2016high}.
 Apart from type-II multiferroic system, a new class of  $A_4B_2O_9$ (where, A = Fe, Co, Mn, and B = Nb, Ta) compounds crystallizing in centrosymmetric crystal structures, exhibit novel linear-ME (LME)effect. Some of the above type of materials exhibit collinear antiferromagnetic structure at the ground state\cite{jana2019low,ding2020successive,khanh2017manipulation,khanh2016magnetoelectric,liu2016magnetodielectric}.  The crystallographic structure of $A_4B_2O_9$  is derived from magnetoelectric corundum-type $Cr_2O_3$, which belongs to the centrosymmetric layered trigonal structure with $P-3c1$  space group\cite{bertaut1961etude}. The layered trigonal crystallographic structures are generally formed by two different types of honeycomb layers stacked alternately along the $c$-axis.  Recently $Mn_4Nb_2O_9$, $Co_4Nb_2O_9$ and $Fe_4Nb_2O_9$ are reported to belong to LME class\cite{jana2019low,fang2014large,fang2015magnetic,maignan2018type}. 

Temperature dependent magnetic susceptibility $\chi(T)$ and neutron diffraction measurements on honeycomb $Fe_4Nb_2O_9$ (FNO) show the antiferromagnetic transition at   $T_N$ = 94K, with a strong magnetic anisotropy suggesting in-plane antiferromagnetic exchange iterations and out of plane ferromagnetic exchange interactions between Fe atoms along $c$-axis \cite{jana2019low,lee2019highly,ding2020successive}. Dielectric permittivity measurements on the same material show two anomalies at 94 and 80 K, due to magnetic transition and trigonal $P-3c1$ to monoclinic $C2/c$ structural phase transition, respectively \cite{jana2019low,ding2020successive,fiebig2005revival,ding2016one}.

Recently it is reported that in LME $Co_4Nb_2O_9$, the lattice distortion causes the change in bond strength leading to anisotropic magneto-dielectric coupling phenomenon \cite{khanh2019anisotropic}. Hence it is important to understand the effect of strain on the physical properties of FNO, as it also belongs to the LME class. FNO contains $3d^{6}$ $Fe^{2+}$ magnetic cations having high spin state (S=2), consistent with the $O^{2-}$ and $Nb^{5+}$ oxidation state. Jahn-Teller (JT) distortion of $Fe^{2+}O_6$ octahedra removes the degeneracy of $t_{2g}$ orbitals and consequently tends to localize electrons via a local lattice distortion \cite{huang2017jahn}. Carrier delocalization can be possible by lifting the JT distortion of the octahedra. In strongly correlated systems charge delocalization are identified from the electron-phonon scattering rate that can be obtained from low-frequency electronic (LFE) contributions and the softening of the high-frequency electronic(HFE) peak from Raman spectroscopy measurements \cite{gupta1996electronic,marrocchelli2007pressure,congeduti2001anomalous,yoon1998raman,
katsufuji1994electronic}. 
The reduction of octahedral distortion (hence the JT coupling) can be possible by applying the external pressure on the system. Therefore studying the physical behaviour of FNO at high pressures will be very interesting, which is lacking as far as the knowledge of the authors.
          
 In this work, we have carried out high-pressure x-ray diffraction (XRD) and Raman scattering experiments on polycrystalline FNO. Detailed high pressure XRD data analysis allows us to explain the change in bond lengths formed by the heavy ions in the honeycomb layers and determine a structural phase transition from ambient trigonal $P-3c1$ to monoclinic $C2/c$ phase at about 8.8-9.3 GPa. The FNO unit cell is found to be more compressible along $c$-axis in comparison to $ab$-plane, indicating competition between two types of exchange interaction at high pressures. Our pressure-dependent Raman scattering data show a change in the slope and a minimum in the FWHM of selected Raman modes around 8.8 GPa. 
The anomalous behaviour of Raman modes in combination of analysis of the structural data indicates a magnetic transition along with a possible change in lattice electrical polarization in the monoclinic phase at high pressures.

 \section{Experimental} 
Powder FNO samples were synthesized through a conventional solid-state synthesis route as described in literature \cite{jana2019low}. We have carried out high-pressure XRD and Raman spectroscopy measurements using a piston-cylinder type diamond anvil cell (DAC) from EasyLab Co. (UK) having 300$\mu$m culet. Polycrystalline FNO samples were crushed into a very fine powder and loaded inside a hole of 120$\mu$m diameter drilled in a pre-indented stainless-steel gasket. The indented thickness of the gasket was about 50$\mu$m. Methanol-ethanol $(4:1)$ mixture was added as a pressure transmitting medium (PTM).
The pressure-dependent XRD measurements at room temperature were carried out at XPRESS beamline in ELETTRA synchrotron source in Trieste, Italy. The wavelength of the monochromatic x-ray beam was 0.4957 $\AA$. The incident high energy x-ray radiation was collimated to about 30 $\mu$m. The diffraction patterns were recorded using $MAR-345$ image plate detector, placed normal to the incident beam. A minute amount of fine silver powder was added along with the sample during loading of DAC. Equation of state of silver was used for the determination of pressure values {\it in situ} \cite{dewaele2008compression}. The sample to the detector distance was calibrated using $LaB_6$. Diffracted patterns were integrated to $2\theta$ vs intensity profile using FIT2D software \cite{hammersley1996two}. The XRD patterns were indexed using CRYSFIRE \cite{shirley2002crysfire} and for the refinement of the unit cell parameter, we used CHECKCELL \cite{laugier2004lmgp}. Lebail and Rietveld refinements of all the XRD patterns were carried out using GSAS software \cite{toby2001expgui}.

Pressure dependent Raman spectroscopy measurements were performed using the confocal micro-Raman system (Monovista from S\&I GmbH). Raman signals were collected in the back scattering geometry. A few ruby chips (approximate sizes of 3-5 $\mu$m) were added along with the sample during loading the DAC for pressure calibration using the ruby fluorescence technique \cite{mao1986calibration}. The sample was excited using Cobalt-samba diode pump 532nm laser source. Raman signals were collected using a infinitely corrected 20X objective lens having a large working distance and a grating of  1500 rulings/mm with a spectral resolution of about 1.2 $cm^{-1}$. For detailed Raman analysis, we have collected the Raman spectra from 100 cm$^{-1}$ to get rid of any effect of the edge filter used to remove the Rayleigh line, which was standardized from the Raman spectra of $Si$ (Supplementary Fig.S1). 
    
\section{Results and discussions} 
We have collected the XRD patterns of polycrystalline FNO at different pressures up to 30 GPa. At ambient conditions, all the diffraction peaks can be well indexed to the trigonal unit cell with $P-3c1$ symmetry. The Rietveld refinement of the ambient pattern is carried out using the initial atom positions obtained from Jana \emph{et~al.}\cite{jana2019low} and shown in Fig.1(a).
The calculated lattice parameters from the best fit are a = 5.23221(13) $\AA$, c = 14.2259(8) $\AA$, and corresponding unit cell volume V= 338(6)$\AA^3$. These are quite close to the literature values \cite{jana2019low}. 
In Supplementary Fig.S2 we have plotted pressure dependence of XRD patterns at selected pressure points.
The XRD patterns of the FNO sample show no significant changes up to 8.8 GPa. Changes in the diffraction patterns are observed with the appearance of splitting of certain Bragg peaks around 9.3 GPa as shown in the Supplementary Fig.S2(a). With further increase in pressure more Bragg peaks start splitting and they are found to be well resolved by about 15.3 GPa (Supplementary Fig.S2(b)). We indexed the XRD pattern at 12.4 GPa, which returned a monoclinic structure with lattice parameters, $a$ = 8.9215(1),  $b$ = 5.1199(4), $c$ = 13.8901(2) $\AA$, $\beta = 91.461(9) ^o$, and corresponding unit cell volume V= 634.26(3) $\AA^3$.
For a better understanding of the crystal structure at high pressures, we have performed group subgroup analysis using Bilbao Crystallographic Server tools. The symmetry-related maximal subgroup of $P-3c1$ is the subgroup $C2/c$ having space group number 15. The XRD pattern at 12.4 GPa is best matched to the above lattice parameters in the C2/c space group. The new structure is very much similar to the antiferromagnetic monoclinic phase obtained by Jana et. al. below 85 K \cite{jana2019low}. 
The Rietveld refinements of the monoclinic phase are carried out by calculating the atom position parameters using group-subgroup analysis using the software Powder Cell \cite{kraus1996powder,kraus2000powdercell}.
The best Rietveld fit of the XRD pattern at 12.4 GPa is shown in Fig.1(b) using new atom positions which are given in Table I. We used the same monoclinic structure to index the XRD patterns from 9.3 GPa, which gave a good match. Therefore, we observe a structural transition at about 9.3 GPa from the trigonal to the monoclinic phase.  No changes are observed upon a further increase in pressure except the broadening of peaks. 

The pressure evolution of lattice parameters are shown in Fig.2(a) for both the phases. The $c$-axis does not show any drastic change across the phase transition pressure. However $a$-axis shows a jump at the transition pressures. The $b$-axis, in case of the monoclinic phase shows a slight upturn at high pressures. 
The unit cell volume obtained after refinement of each XRD pattern at all pressure points (in both the phases) fitted to the 3rd order BM-EOS function \cite{angel2014eosfit7c} and is shown in Fig.2(b):
\begin{equation}
	P(V)=\frac{3 B_{0}}{2}\left[\left(\frac{V_{0}}{V}\right)^{7 / 3}-\left(\frac{V_{0}}{V}\right)^{5 / 3}\right] \times\left\{1+\frac{3}{4}\left(B^{\prime}-4\right)\left[\left(\frac{V_{0}}{V}\right)^{2 / 3}-1\right]\right\}
\end{equation}
where $B_0$ and $B'$ are the bulk modulus and the first-order pressure derivative of bulk modulus, respectively. $P$ is the pressure, $V_0$ is the ambient unit cell volume. The obtained EOS parameters are: for the trigonal phase $B_0$ = 155 $\pm$ 11 GPa, $B'$ = 5.4; and for monoclinic phase $B_0$ = 167 $\pm$ 5 GPa, $B'$ = $3.7$. Our results indicate a decrease in compressibility with the structural transition.
  
It is interesting to see the unit cell arrangement of FNO. The trigonal phase consists of layers of metal-oxygen polyhedra along the $c$-axis. There are two layers formed by edge shared $NbO_6$ and $Fe2O_6$ octahedra separated by a single layer formed by edge sharing $Fe1O_6$ octahedra along $c$-axis. With the application of pressure, the distortion index and the quadratic elongation of both the $Fe1-O_6$ and $Fe2-O_6$ octahedra decrease till about 8.8 GPa and then suddenly jumps in the monoclinic phase. In contrast, both the distortion index and the quadratic elongation of $Nb-O_6$ octahedra reach their respective maximum values at the phase boundary of trigonal to monoclinic phases as shown in the Fig.3(a) and Fig.3(b). Since there is a minimal change in volume across the structural transition (from 160.5 \AA$^3$ at 8.8 GPa to 160.2 \AA$^3$ at 9.3 GPa), the large distortion of $Nb-O_6$ octahedra may be driving the structural transition. The effect of small volume change can be seen from the starting of splitting of certain Bragg peaks at 9.3 GPa, which continue until they are well resolved at 12.4 GPa.  Another important part is that the ambient trigonal phase undergoes a magnetic transition to an antiferromagnetic state below 95 K \cite{maignan2018fe,jana2019low,ding2020successive}, in which the spins of Fe atoms are aligned ferromagnetically along $c$-axis and antiferromagnetically in the $ab$-plane. Therefore we monitored the changes in the separation of $Fe1-Fe1$ atoms and $Fe1-Fe2$ atoms in the $ab$-plane and along the $c$-axis, respectively in Fig.3(c). We find that for both the cases the separation decreases and the relative change for $Fe1-Fe2$ atoms along $c$-axis is more compared to that for $Fe1-Fe1$ in the $ab$-plane. However, from this information alone it is not possible to suggest the changes in magnetic behaviour of FNO at about 8.8 GPa, as the net magnetic structure depends on the competition among several  exchange interactions present in the solid.

For complementary investigation of FNO under pressure, we have carried out the Raman spectroscopy studies on the same. We shall first discuss the ambient Raman spectrum as shown in Fig.4(a). The spectrum shows 7 Raman peaks with a broad continuum, the intensity of which increases below 200 cm$^{-1}$ and shows a broad peak in the range 500 - 900 cm$^{-1}$. 
FNO crystallizes in the trigonal crystal structure with the space group $P-3c1$, having two formula units and it has 30 atoms per primitive cell which leads to 57 optical phonon and 3 acoustic phonon modes in the Brillouin zone center \cite{chen2019attributions}. Group theory predicts the following representation of all the vibrational modes in the zone center boundary: $6A_{1g}+9A_{1u}+6A_{2g}+9A_{2u}+12E_{g}+18E_{u}$, in which all the $A_{1g}$ and $E_{g}$ vibrational modes are Raman active. Hence there are 18 Raman active modes but experimentally we see only seven well defined modes. 
The well defined modes are labeled as: $\nu_1=220 cm^{-1}, \nu_2=282 cm^{-1}, \nu_3=396 cm^{-1}, \nu_4=490 cm^{-1}, \nu_5=598 cm^{-1}, \nu_6=650 cm^{-1}$,and $\nu_7=850 cm^{-1}$. The ambient Raman spectrum of FNO is in good agreement  with the reported spectra \cite{chen2019attributions}. Following the literature we assign $\nu_1$, $\nu_2$ and $\nu_5$ to $E_g$ modes; $\nu_3$ and $\nu_7$ as $A_{1g}$ modes \cite{chen2019attributions,rodrigues2017raman}. $\nu_3=396 cm^{-1}$ and $\nu_7=850 cm^{-1}$ are internal vibrational modes due to the stretching and breathing motions of $NbO_6$ octahedra, respectively \cite{rodrigues2017raman,rodrigues2017ordering,rodrigues2016structural}.

Next we shall discuss the general observation of the pressure evolution of Raman modes, which are shown in Fig.4(b). All the mode frequencies show a blue shift with increasing pressure as expected, due to compression in volume. We observe that  a new mode ($\nu^{'}_1$)appears at a low pressure of about 0.9 GPa. This new mode becomes stronger with pressure. The mode $\nu_2$ splits into two modes, $\nu^{'}_2$ and $\nu^{''}_2$. With increase in pressure the intensity of $\nu^{''}_2$ increases and becomes comparable to $\nu^{'}_2$. 
The splitting of the two low frequency modes at such low pressure value without any structural transition is quite an interesting phenomenon. May be there are some microscopic changes in  the FNO unit cell, which can give rise to a local ordering effect resulting in the appearance of new peaks in Raman spectrum. It can be noted here that a sudden decrease in $Nb-O_6$ octahedra distortion is observed from room pressure value to 0.8 GPa (see Fig.3), beyond which it increases sharply. In fact as we discussed earlier, during the pressure induced structural transition, only a large change in the distortion of $Nb-O_6$ octahedra is observed. Therefore in all possibility, the emergence and splitting of Raman modes are probably driven by change in local structural order due to change in $Nb-O_6$ octahedral distortion.
Since we do not find any increase in the number of Raman modes above the phase transition pressure to the monoclinic structure, we believe that the $Nb-O_6$ octahedra distortion induces a formation of local monoclinic structure at the microscopic level. This is probably missed by the XRD measurements since it gives an average crystal structure. Above 9.3 GPa the monoclinic phase attains a significant proportion to be detected by the XRD.
The low frequency diffusive continuum becomes gradually flat with pressure and almost disappears above about 6.7 GPa (Fig. 4(b)). Similarly, the broad continuum in the range 500 - 900 cm$^{-1}$ decreases above about 8.8 GPa. The strong pressure dependence shows this feature to be intrinsic to the sample in origin. Similar features are seen in other strongly correlated oxide insulators showing magnetic anomalies \cite{congeduti2001anomalous,marrocchelli2007pressure,yoon1998raman,ulrich2015spin,gupta1996electronic,iliev2001raman}.

Further detailed analysis of the pressure evolution of Raman spectra are carried out by multiplying each spectrum using the Bose-Einstein thermal factor. The Raman-scattering response of FNO in the 100-950 $cm^{-1}$ frequency range can be well fitted (as shown in Fig.4(a)) with the spectral response,\cite{yoon1998raman}

\begin{equation}
S(\nu)= S_{ph}(\nu)+S_{el}(\nu),
\end{equation}

\noindent where 

\begin{equation}
S_{ph}(\nu)=[1+n(\nu)] \sum_{1}^{n}\frac{A_i\nu\Gamma_i}{(\nu^2-\nu_i^2)^2+\nu^2\Gamma_i^2}
\end{equation}

\noindent represents $n$ phonon peaks ($n$ = 7 for ambient and $n$ = 9 above 0.9 GPa) with associated peak frequency: $\nu_i$, amplitude: $A_i$ and line width: $\Gamma_i$.

\begin{equation} 
S_{el}(\nu)=[1+n(\nu)][\frac{A\nu/\tau}{\nu^2+1/\tau^2}+\frac{A_{el}\nu\Gamma_{el}}{(\nu^2-\nu_{el}^2)^2+\nu^2\Gamma_{el}^2}],  
\end{equation}

\noindent where the first right hand side term represents low frequency electronic (LFE) response due to diffusive hopping of the carriers with a scattering rate $1/\tau$ and the second term represents high frequency electronic (HFE) contribution near $\nu_{el}\sim 550-900 cm^{-1}$. The term $[1+n(\nu)]$ represents the Bose-Einstein thermal factor. Our data fits extremely well to the above equations (Fig.4(a)).

Fig.5(a) shows the pressure evolution of significant Raman modes. All the modes show linear behaviour with respect to pressure with changes in the slope around 8.8 GPa. XRD studies have revealed a structural transition in FNO from trigonal to the monoclinic phase in the range 8.8 - 9.3 GPa. Therefore the change in slope can be attributed to the above phase transition. The slope values of the mode frequencies with  respect to pressure decrease significantly above the transition pressure. The decrease in slopes of the pressure evolution of Raman modes in the monoclinic phase with respect to those in the trigonal phase can be related to the decrease in compressibility. However, it may be noted that the slopes change by around 50$\%$ where as the compressibility changes only by 8$\%$, possibly other factors like, a change in physical properties along with the structure are responsible. The $Nb-O_6$ octahedral breathing mode, $\nu_7$ shows the maximum change in frequency with pressure. It can be related to the large distortion of $Nb-O_6$ octahedra with pressure (Fig.3). The $Fe-O_6$ octahedra show a gradual decrease in their distortion till the transition pressure. Therefore as we discussed earlier, the structural transition can also be viewed as to be driven by increased distortion in the $Nb-O_6$ octahedra, which becomes more ordered in the monoclinic phase.
Interestingly the FWHM of $\nu_1^{\prime}$, $\nu{'}_2$, and $\nu_3$ show a minimum at 8.8 GPa (see Fig.5(b)). The FWHM of $\nu_1$ keep on decreasing till the highest pressure studied. The width of other modes keep broadening with pressure. The FWHM of a Raman mode is associated with the life time of the phonon mode. The anharmonic phonon-phonon scattering events increase the FWHM and hence decrease the lifetime. Therefore, decrease in FWHM of Raman mode indicates a decrease in anharmonic phonon interactions. Application of pressure gradually induce a strain in the lattice, which is supposed to reduce the lifetime of phonons. A minimum in phonon-modes FWHM can indicate a possibility towards a strong phonon-electron interaction, which causes decrease in the life time of phonons after an electronic phase transition \cite{jana2016high, gupta2017raman, saha2018structural,bera2013sharp}. 
 To look into the Raman scattering events we calculated the normalized intensity of the Raman modes. We have carried out the vector normalization  followed by Gautam et al. \cite {gautam2015review}. In the vector normalization, first of all the "norm" of the spectrum is calculated. Then to obtain the normalized Raman spectrum, each of the Raman intensities corresponding to a wavenumber is divided by the "norm" of that spectrum. Ideally the intensity of the Raman modes are supposed to reduce under pressure due to the induce lattice strain. Interestingly, the normalized integrated intensity of the $\nu_1^{\prime}$, $\nu_2^{\prime\prime}$ and $\nu_3$ Raman modes show a maximum at about 8 GPa as shown in the Fig.6. The sharpening of these modes along with the increase in the intensity till about 8 GPa shows reduction in anharmonic phonon scattering events. These results indicate an electronic transition along with the structural transition as described earlier. 
At this point it is important to point out that low temperature studies on FNO has revealed an antiferromagnetic transition at about 95 K followed by a structural transition to a monoclinic phase at about 85 K, which remained in antiferromagnetic state \cite{jana2019low}. Low temperature Raman spectroscopic studies on a sister material $Mn_4Nb_2O_9$ do not show any anomalous behaviour in its Raman modes across its antiferromagnetic transition temperature of about 110-130 K \cite{chen2019attributions}.

Let us now look into the behaviour of the electronic Raman scattering response by analyzing the response of LFE and HFE contributions with pressure. The LFE scattering rate showing an almost exponential decrease with the maximum drop till about 8 GPa (Fig.7(a)). The XRD analysis confirms the gradual decrease of the distortion of both the $FeO_6$ octahedra with increasing pressure in the trigonal phase. Decrease in LFE scattering rate also indicates the sharp reduction of lattice distortion with increasing pressure. Therefore one can assume that the Jahn-Teller distortion of the octahedra decreases gradually, which consequently increases the charge delocalization leading to a possibility of insulator to metal transition.
 On the other hand, the peak value of the high frequency inelastic electronic contribution hardens up to about 8 GPa and its intensity decreases by an order of magnitude up to the about 14 GPa (Fig.7(b) and Fig.7(c)). The HFE contribution is suggested to be associated with the local polaron formations due to lattice distortions \cite{yoon1998raman}. The peak value of HFE contribution is attributed to the ionization energy of the polarons. The maximum peak value of HFE contribution is approximately equal to about 91 $meV$, which is much smaller than the polaron formation energy \cite{jaime1996high} as well as estimates obtained for band structure due to Jahn-Teller splitting \cite{satpathy1996electronic} and also estimates obtained for perovskite manganese oxides \cite{yoon1998raman}. In addition, reduction of HFE intensity shows that there is a decrease in scattering cross section of phonons with excited electrons leading to decrease in density of states at the Fermi energy \cite{gupta1996electronic}. Several other strongly correlated systems, those undergo paramagnetic insulator phase to ferro-/antiferro-magnetic metallic transition at low temperatures, have shown hardening of HFE contribution in both intensity as well as its peak position and also decrease in LFE contribution up to the transition temperature \cite{billinge1996direct,girardot2008raman,yoon1998raman}. In a high pressure Raman study on a manganese oxide perovskite system both the LFE and HFE contributions did not show the expected behaviour of insulator to metal transition and the authors predicted a new un-characterized phase at high pressures \cite{congeduti2001anomalous}. Therefore, considering our LFE and HFE contributions with pressure, we can exclude an insulator to metal transition in our system.
 
	Though across the low temperature phase transition the distortion index of $Nb-O_6$ octahedra remains almost same(decreases by only about 3\%), both the $Fe-O_6$ octahedra also show a significant increase in distortion index \cite{jana2019low}. In the present work, the distortion index of both the $Fe-O_6$ octahedra reduces significantly up to the transition pressure but after the transition pressure the distortion shows a sudden increase (for $Fe1-O_6$ octahedra it is 31\% and for $Fe2-O_6$ octahedra it is about 18\%). 
  
	 The evidence of the increasing distortion index of both the $Fe-O_6$ octahedra indicate a JT distortion in the monoclinic phase which might be responsible for the emergence of the magnetic state. However, it is difficult for us to comment on the exact nature of magnetic transition in the absence of any magnetization measurements at such high pressure. We can only speculate that the magnetic nature of this high-pressure monoclinic phase will be similar to that observed in case of low temperature monoclinic phase which is shown to be antiferromagnetic.
  
   In a low temperature study on FNO, Maignan and Martin \cite{maignan2018fe} observed a large peak in the dielectric constant as well as large polarization current (which flipped its sign with reversal of electric field) just below the antiferromagnetic transition temperature. They attributed these observations to the large magnetoelectric effect of FNO in its antiferromagnetic state. It is well known that the normalized Raman mode intensity also depends on the polarization of that particular mode. Therefore, we expect that a change in ferroelectric polarization in the sample may show up in the behaviour of normalized Raman mode intensity. In Fig.6 we have plotted the pressure evolution of the normalized integrated intensity of $\nu_1$,  $\nu_1^{\prime}$, $\nu_2^{\prime}$, $\nu_2^{\prime\prime}$ and $\nu_3$ modes. The $\nu_1^{\prime}$, $\nu_2^{\prime\prime}$ and $\nu_3$ modes show a maximum in the intensity around 8 GPa, which also show a minimum in their FWHM at the same pressure range. The $\nu_1$ and $\nu_2^{\prime}$ mode intensities are found to keep increasing till the highest pressure value with an anomaly at about 15 GPa. The increase in the normalized integrated intensity of all the above modes with pressure may be attributed to a large change in polarization of the sample. Earlier we have discussed about development of local microscopic order in the sample as a minor monoclinic phase above 0.9 GPa driven by the $Nb-O_6$ octahedra distortion. This most probably is induced by a gradient in the lattice strain, and creates local phases with ferroelectric order. With pressure, the change in local polarization is refelcted in the increase of the Raman mode intensity. Similar strain induced nanoscale topological structures with with distinct ferroelectric phase have been observed by Yudin et. al.  {\cite {yudin2021}}.
 
 From our high pressure XRD study we speculate a magnetic transition along with the structural transition about 8 GPa pressure. There may be a possibility that the change in magnetic structure may induce the anomalies seen in the pressure behavior of the Raman modes due to spin-phonon interaction. But the effect on Raman mode intensity across any magnetic phase transition in the absence of any applied magnetic field has not been reported to the best of our knowledge. It has been observed that a very high magnetic field is required to produce any significant change across the magnetic transition in the Raman mode intensity \cite{sooryakumar1981raman, ji2016giant, mccreary2020distinct, li2000interface}. Since we have excluded any insulator to metal transition in FNO at high pressure one can also assume that the spin induced large change in polarization may give rise to the interesting behaviour of LFE and HFE contributions to the Raman spectra. Hence we may conclude that FNO also exhibits a manetoelectric behaviour in its high pressure monoclinic phase.

\section{Conclusion} 
The phase transitions in $Fe_4Nb_2O_9$ are investigated in detail at high pressures by means of  X-ray diffraction and Raman Spectroscopy measurements. Our XRD  results show a structural transition from ambient trigonal $P-3c1$ to monoclinic $C2/c$ at high pressures. There is a large increase in both the quadratic elongation and distortion index of $Nb-O_6$ octahedra up to about 8.8 GPa followed by sudden decrease in the monoclinic phase. In contrast, both the $Fe-O_6$ octahedra get more ordered with pressure in the trigonal phase. All the prominent Raman modes show a slope change in their linear behaviour at the structural transition pressure. Some of the Raman modes show a minimum in their FWHM at the same pressure indicating an electronic transition. A decrease the diffusive scattering rate is observed in the LFE contribution to the Raman spectra indicating a charge delocalization. However, observed decrease in the amplitude of the HFE contribution and its peak frequency excludes an insulator to metal transition. The intensity of the Raman modes showing minimum in their FWHM, shows a large peak at about 8.8 GPa, indicating a large change in ferroelectric polarization in FNO at the phase transition pressure. Our results show the sample may transform to a magnetoelectric  antiferromagnetic state in its monoclinic phase.

\begin{acknowledgments}
The authors gratefully acknowledge the financial support from the Department of Science and Technology, Government of India to visit XPRESS beamline in the ELETTRA Synchrotron light source under the Indo-Italian Executive Programme of Scientific and Technological Cooperation. MS gratefully acknowledge the CSIR, Govt. of India for the financial support to carry out PhD work.	 

\end{acknowledgments}

\noindent
{\bf {Author Declarations}} 

\noindent All authors have equal contribution. All authors reviewed the manuscript.

\noindent
{\bf {Conflict of interest:}} The authors declare no competing no conflict of interest.

\noindent
{\bf { Data Availability}}: The data that supports the findings of this study are available within the article and its supplementary material.

\pagebreak
\begin{table}[h]
\centering
\caption{Atomic coordinates and isotropic displacement parameters ($\AA^2$) of $Fe_4Nb_2O_9$ refined from XRD data at 12.4 GPa in space group No. 15 ($C2/c, Z=4$). Lattice parameters are $a = 8.9215(1)\AA$, $b = 5.1199(4)\AA$, $c = 13.8901(2)\AA$ $\beta = 91.461(9)^o$ and $V=634.26(3)\AA^3$. The goodness of fit parameters $R_{wp}=3.51$ \% \& $R_p=2.74$ \% } 

\begin{tabular}{|c|c|c|c|c|c|}

\hline
Atom & Site & $x$ & $y$ & $z$ & $U_eq$ \\ 
\hline
Nb & $8f$ & 0.000000 & 0.000000 & 0.160(3) & 0.0095(4) \\
\hline
Fe1 & $8f$ & 0.166(5) & 0.5000 & 0.1996(7) & 0.00532(2) \\
\hline
Fe2 & $8f$ & 0.166(6) & 0.5000 & 0.9678(8) & 0.0095(1) \\
\hline
$O1_1$ & $8f$ & 0.0126(8) & 0.6719(7) & 0.0889(8) & 0.0097(13) \\
\hline
$O1_2$ & $8f$ & 0.15767(5) & 0.18304(5) & 0.0889(8) & 0.0097(13) \\
\hline
$O1_3$ & $8f$ & 0.3296(4) & 0.6449(9) &  0.0889(8) & 0.0097(13) \\
\hline
$O2_1$ & $4e$ & 0.000000 & 0.287000  & 0.250000 & 0.014(3) \\
\hline
$O2_2$ & $8f$ & 0.356(5) & 0.356(5) & 0.250000 & 0.01400(3) \\
\hline
\end{tabular}
\end{table}

\begin{figure}[htb]
    \begin{center}
    \includegraphics[height=7in,width=7in]{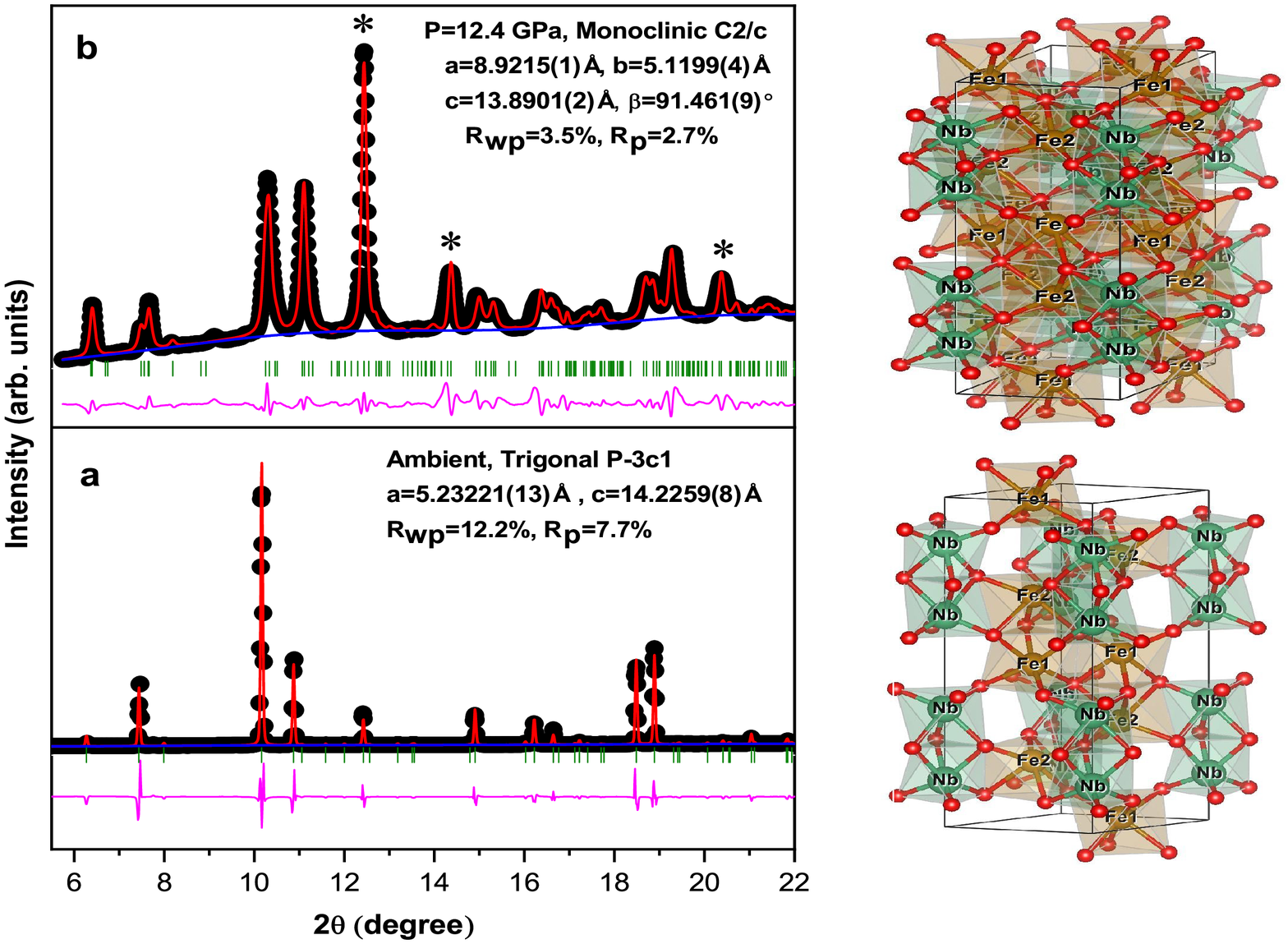}
    \caption{Rietveld refined XRD patterns of polycrystalline $Fe_{4}Nb_{2}O_{9}$ at (a) ambient pressure (trigonal phase in space group $P-3c1$) and (b) 12.4 GPa, after transition (monoclinic phase in space group $C2/c$). The bold black circle indicates observed data points, the fitted values are represented by the red line over the data points, and the pink line below denotes the difference between the observed and fitted intensities. The pressure marker silver peaks are indicated by '*'. The right side of the XRD pattern show the arrangements of the metal oxygen polyhedra with the lower one belonging to the trigonal phase and the upper one to the monoclinic phase.}
    \end{center}
		\end{figure}

\begin{figure}
    \centering
    \includegraphics[scale=0.8]{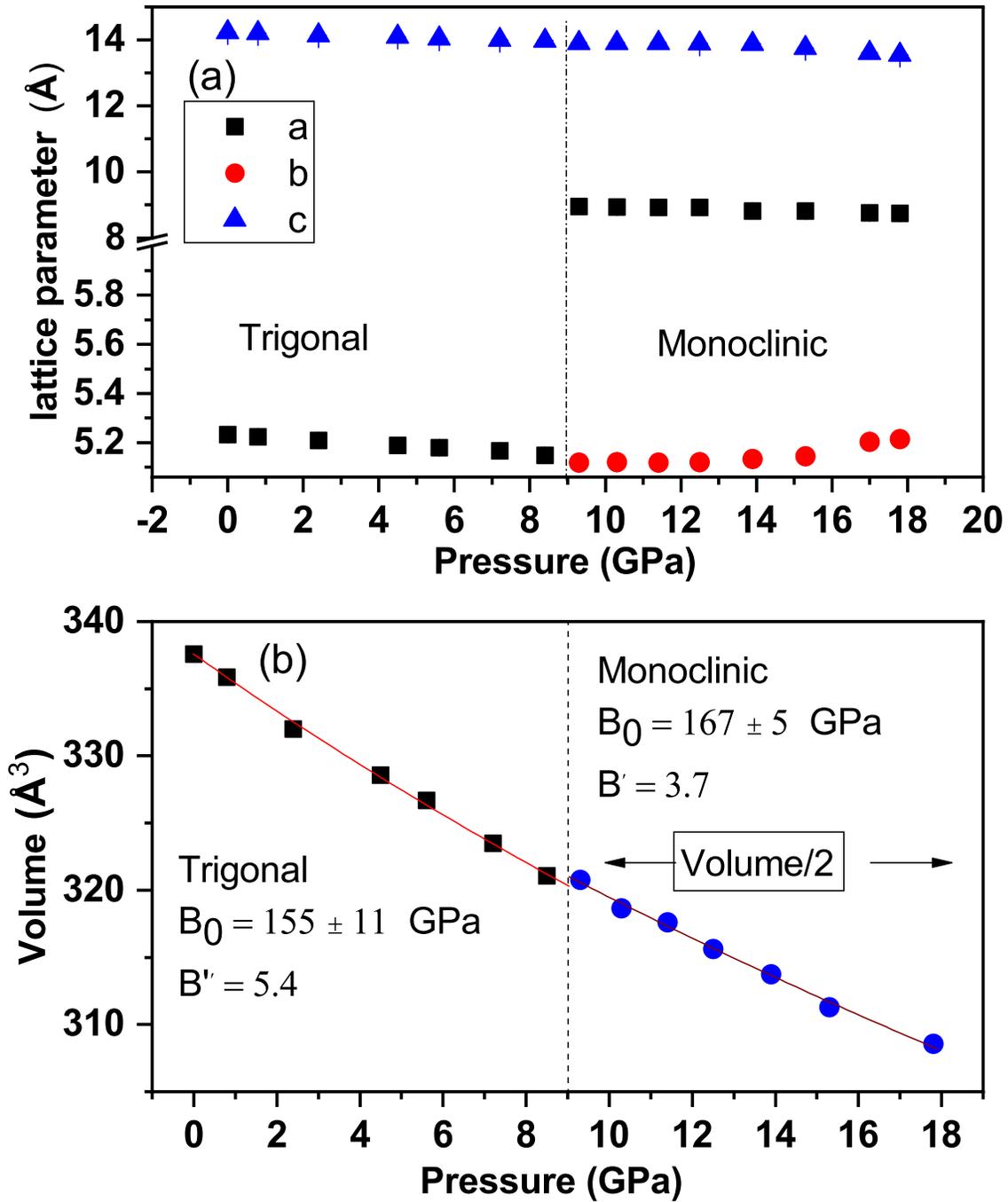}
    \caption{Pressure evolution of (a) unit cell lattice parameters and (b) unit cell volume of polycrystalline $Fe_{4}Nb_{2}O_{9}$. The lines passing through the volume is the Birch-Murnaghan EOS fit to the volume data.}
		\end{figure}

\begin{figure}
    \centering
    \includegraphics[scale=0.6]{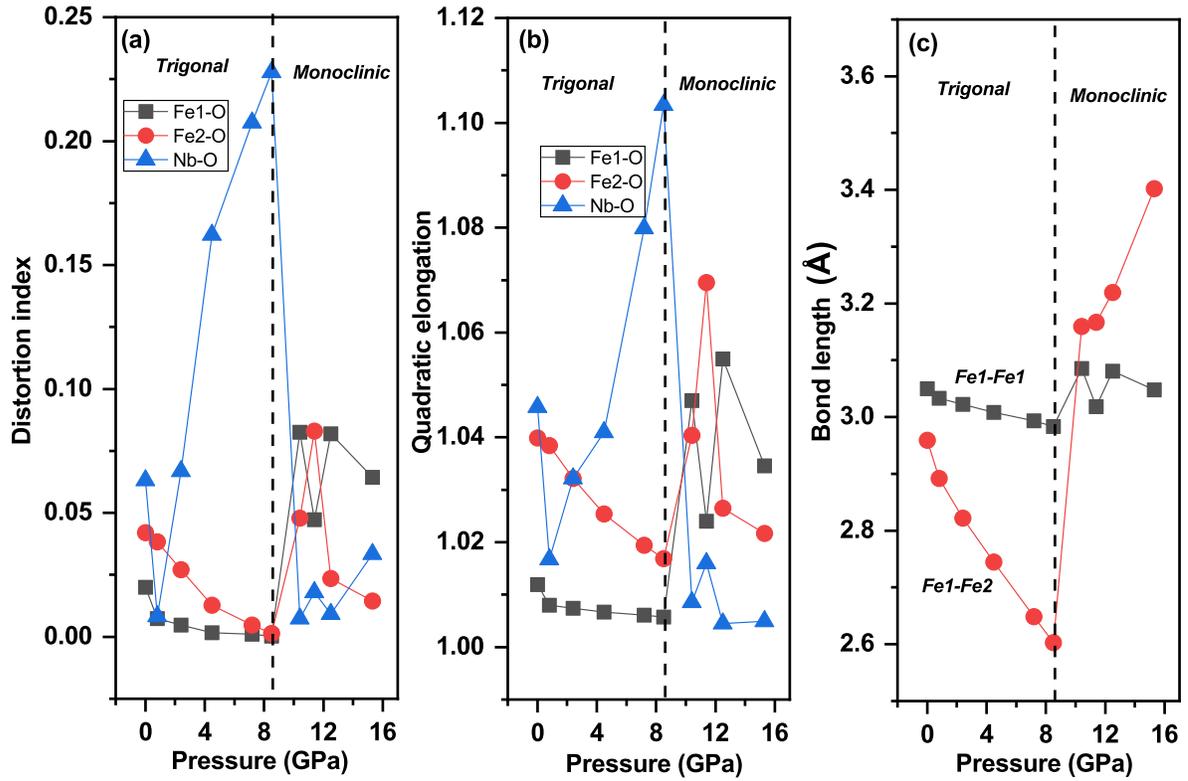}
    \caption{(a) Distortion index and (b) quadratic elongation of metal-oxygen octahedra at different pressures. (c) Distance between $Fe1-Fe2$ atom along $c$-axis and $Fe1-Fe1$ atoms in the $ab$-plane at various pressures The vertical dashed line marks the pressure value at the trigonal to monoclinic phase transition.}
		\end{figure} 

\begin{figure}
    \centering
    \includegraphics[scale=0.8]{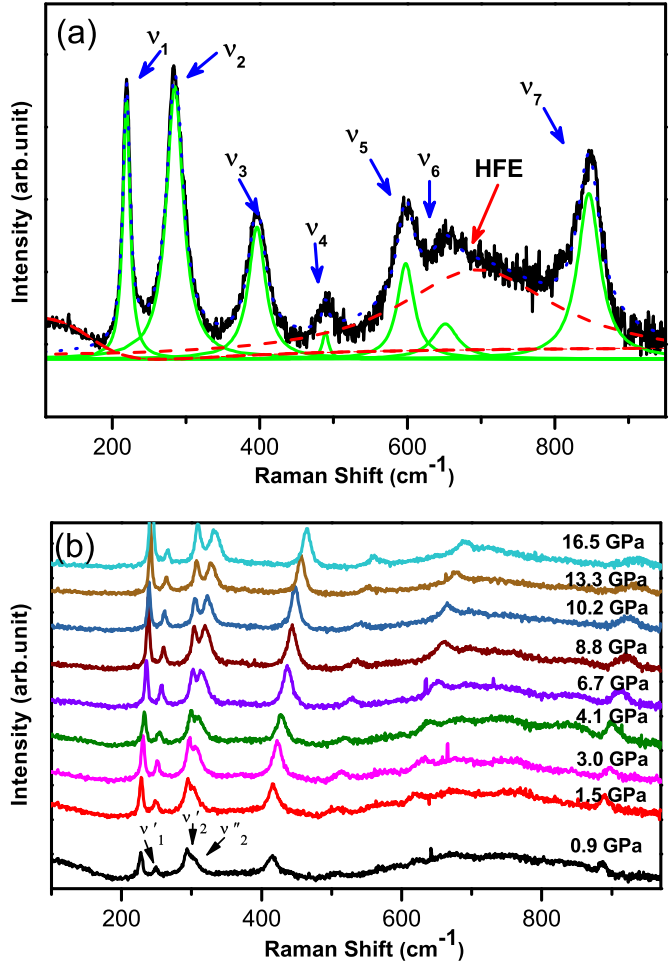}
		
		\vspace{-50 mm}
		
    \caption{(a) The deconvolution of the ambient Raman spectrum along with its fit using the Eq. (b) Raman spectra at selected pressure points. The new mode $\nu_1^\prime$ and the split modes $\nu_2^\prime$ and $\nu_2^{\prime\prime}$ are marked. }
		\end{figure}
		
\begin{figure}[!ht]
    \centering
    \includegraphics[scale=0.6]{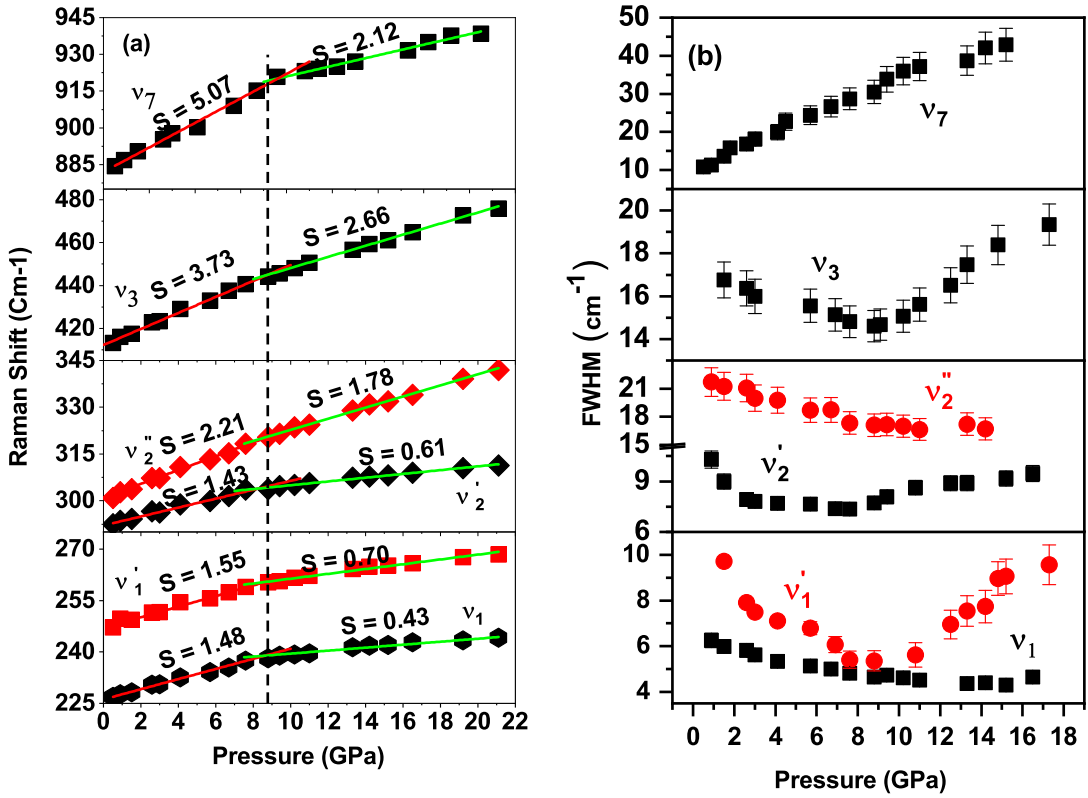}
    \caption{(a) Pressure evolution of significant Raman modes showing changes in the slope in their linear in pressure behaviour at 8.8 GPa. Lines passing through the data points are the linear fit to the data. (b) Pressure evolution of FWHM of Raman modes. }
		\end{figure}

	\begin{figure}
    \centering
    \includegraphics[scale=0.8]{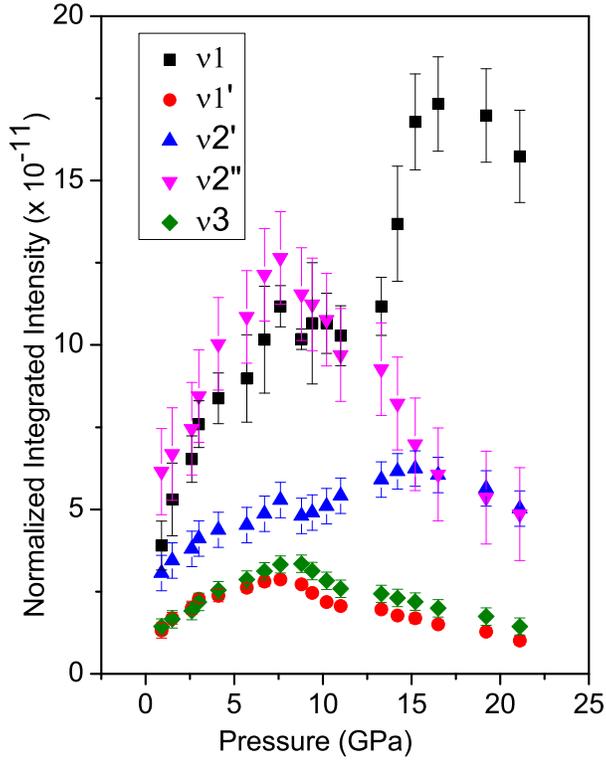}
    \caption{Pressure dependence of the normalized integrated intensity of Raman modes. The $\nu_1^\prime$, $\nu_2^{\prime\prime}$ and $\nu_3$ Raman mode intensities showing a maximum at about 8 GPa pressure. The intensity of the modes, $\nu_1$ and $\nu_2^{\prime}$ increasing with pressure with an anomaly at about 8 GPa.}
		\end{figure}
	
	\begin{figure}
    \centering
    \includegraphics[scale=0.8]{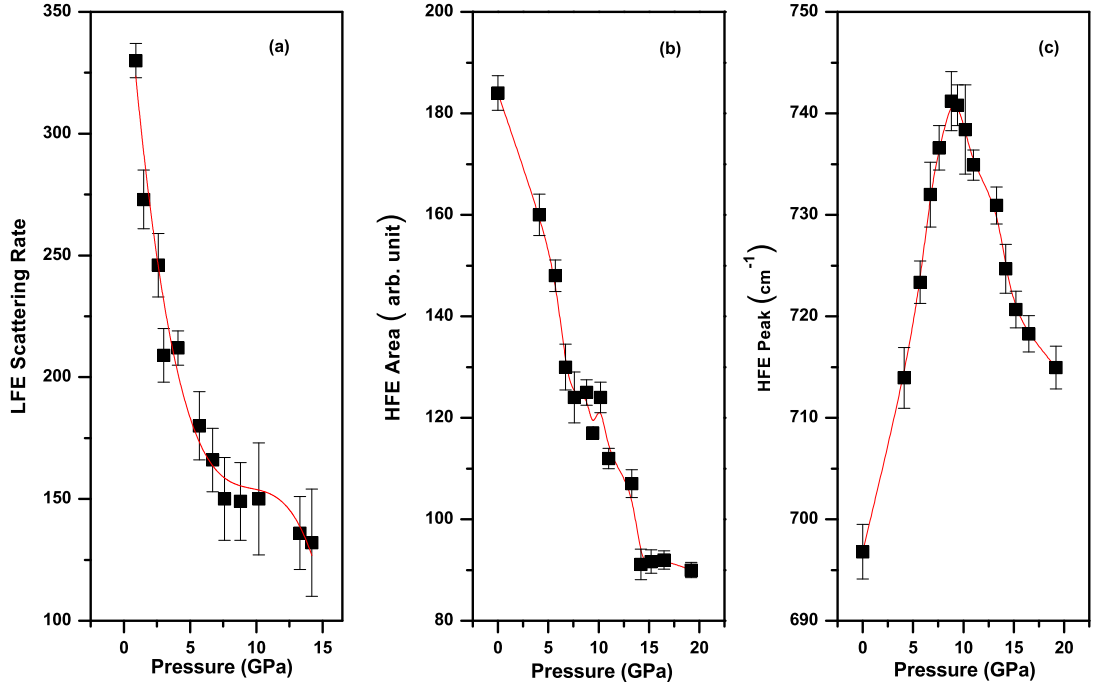}
    \caption{(a) Diffusive carrier scattering rate of the low frequency electronic contribution (LFE) to the Raman spectra at different pressures. (b) Normalized integrated intensity and (c) peak frequency of the high frequency electronic contribution (HFE) to the Raman spectra at different pressures.}
		\end{figure}
		
\end{document}


	
	\title{Supplementary: Pressure driven re-entrant magnetoelectric transition in honeycomb  $Fe_{4}Nb_{2}O_{9}$}
	
	\author{Mrinmay Sahu}
	\affiliation {National Centre for High Pressure Studies, Department of Physical Sciences, Indian Institute of Science Education and Research Kolkata, Mohanpur Campus, Mohanpur – 741246, Nadia, West Bengal, India.}
	
	\author{Bishnupada Ghosh}
	\affiliation {National Centre for High Pressure Studies, Department of Physical Sciences, Indian Institute of Science Education and Research Kolkata, Mohanpur Campus, Mohanpur – 741246, Nadia, West Bengal, India.}
	
	\author{Rajesh Jana}
	\affiliation {Beijing National Laboratory for Condensed Matter Physics and Institute of Physics, Chinese Academy of Sciences, Beijing 100190, China}
	
	\author{Jinguang Cheng}
	\affiliation {Beijing National Laboratory for Condensed Matter Physics and Institute of Physics, Chinese Academy of Sciences, Beijing 100190, China}
	
	\author{Goutam Dev Mukherjee}
	\affiliation {National Centre for High Pressure Studies, Department of Physical Sciences, Indian Institute of Science Education and Research Kolkata, Mohanpur Campus, Mohanpur – 741246, Nadia, West Bengal, India.}
	\email [Corresponding Author:]{ goutamdev@iiserkol.ac.in}

	\date{\today}

	\maketitle
	
	\begin{figure}[htbp]
		\centering
		\includegraphics[scale=0.6]{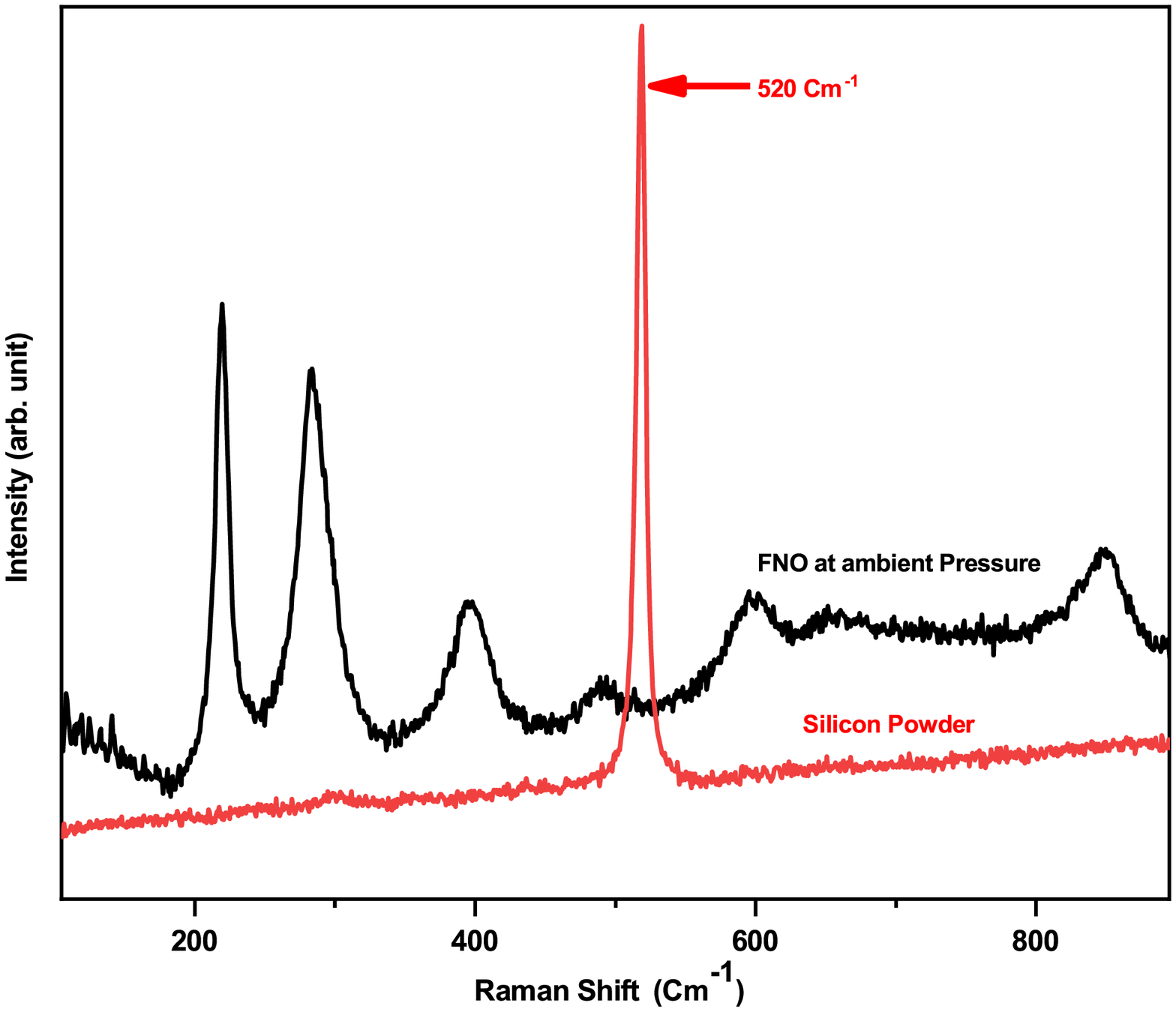}
		\caption{\label{Fig.1} Raman spectrum of FNO and the silicon powder. In the low frequency region ($< 200 Cm^{-1}$), a gradual increase in intensity of the Raman signal of FNO with respect to the silicon powder is clearly seen.}
	\end{figure}

	\begin{figure}[htbp]
	\centering
	\includegraphics[scale=0.6]{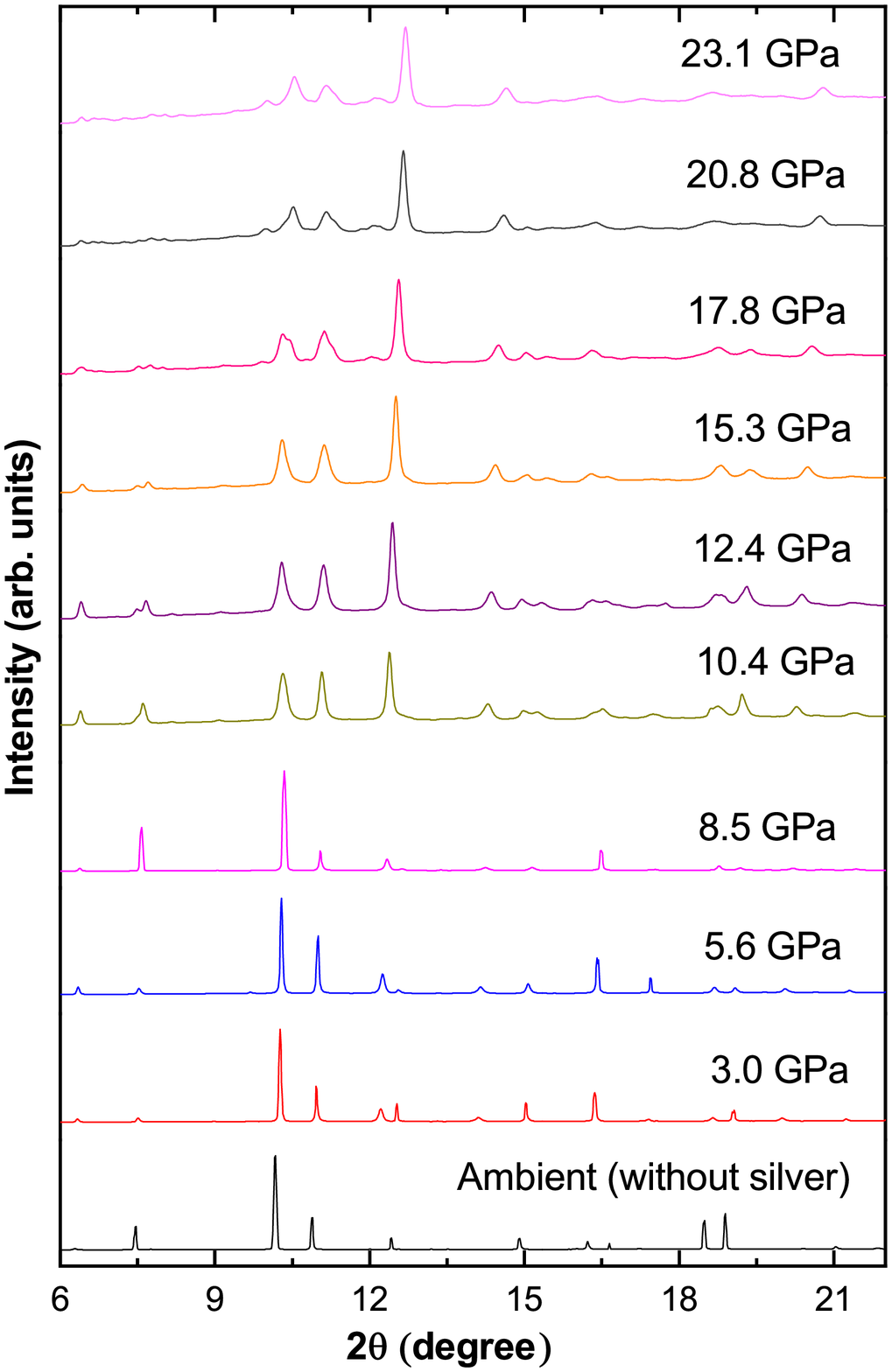}
	\caption{\label{Fig.2}Pressure dependent XRD patterns at selected pressure points.}
	\end{figure}

\begin{figure}[htbp]
		\centering
	\includegraphics[scale=0.6]{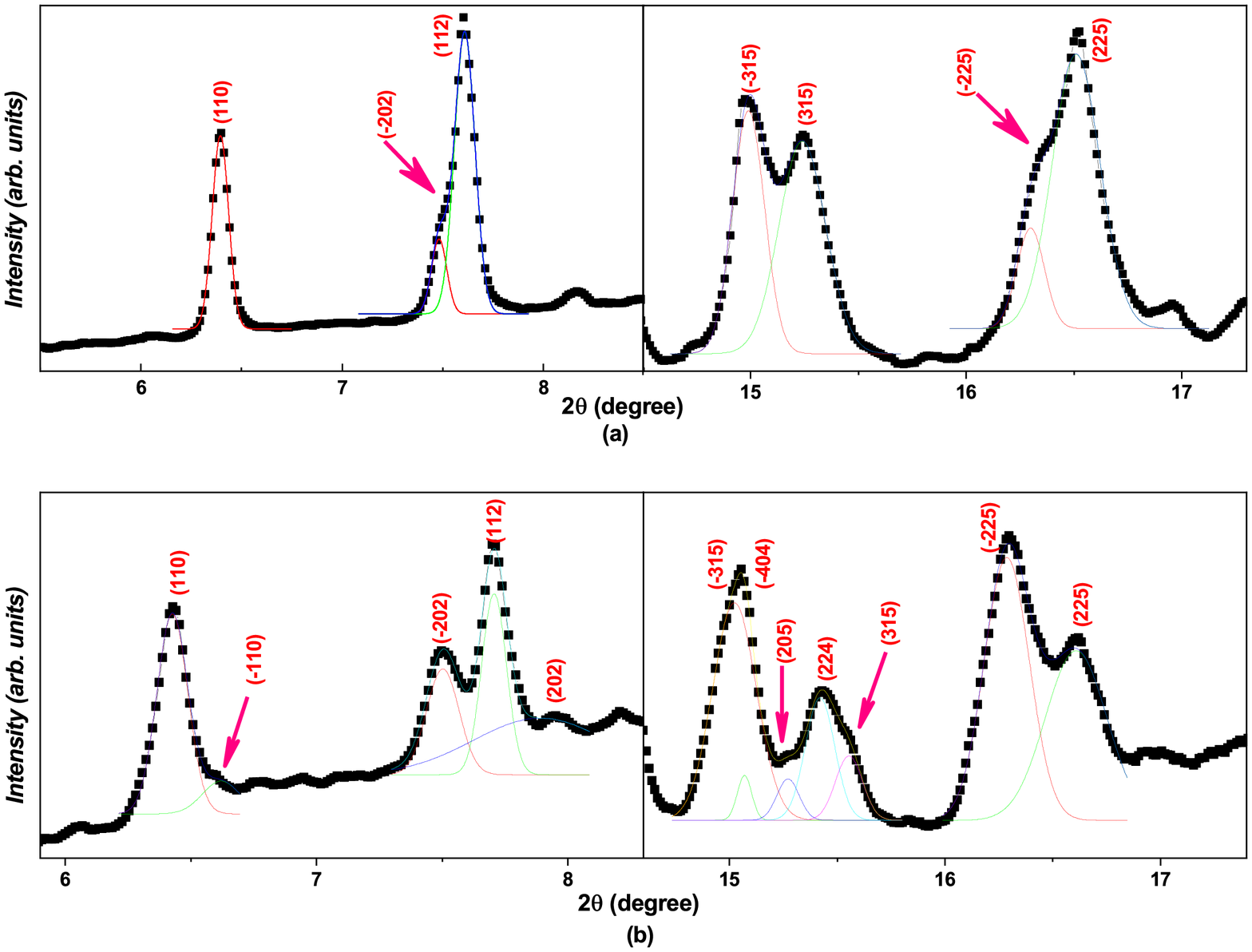}
	\caption{\label{Fig.3}Fitting of Bragg peaks of FNO at 9.3 and and 15.3 GPa. (a) At 9.3 GPa Bragg peak (012) splits into two Bragg peaks (-202) and (112), (204) splits into two, (-315) and (315) and the Bragg peak (116) splits into (-225) and (225) after trigonal to monoclinic phase transition. (b) At 15.3 GPa the low angle Bragg peak (100) in the trigonal phase splits into (110), (-110) of the monoclinic phase. Other peaks also splits into more number of well resolved Bragg peaks in the monoclinic phase.}
\end{figure}